\documentclass[a4paper]{article}

\usepackage{INTERSPEECH2021}
\usepackage{multirow}
\usepackage{url}
\usepackage{colortbl}
\usepackage{hhline}
\usepackage[table,dvipsnames]{xcolor}
\usepackage{nicematrix}

\makeatletter
\def\hlinewd#1{%
\noalign{\ifnum0=`}\fi\hrule \@height #1 \futurelet
\reserved@a\@xhline}
\makeatother

\title{Stabilizing Label Assignment for Speech Separation by Self-supervised Pre-training}
\name{Sung-Feng Huang$^1$, Shun-Po Chuang$^1$, Da-Rong Liu$^1$, Yi-Chen Chen$^1$, \\
Gene-Ping Yang$^2$, Hung-yi Lee$^1$}
\address{
    $^1$National Taiwan University, Taiwan\\ 
    $^2$University of Edinburgh, UK}
\email{
    \{f06942045,f04942141,f07942148,f06942069\}@ntu.edu.tw,\\
    s2064029@ed.ac.uk, hungyilee@ntu.edu.tw}

\begin{document}

\maketitle
\begin{abstract}
Speech separation has been well developed, with the very successful permutation invariant training (PIT) approach, although the frequent label assignment switching happening during PIT training remains to be a problem when better convergence speed and achievable performance are desired.
In this paper, we propose to perform self-supervised pre-training to stabilize the label assignment in training the speech separation model.
Experiments over several types of self-supervised approaches, several typical speech separation models and two different datasets showed that very good improvements are achievable if a proper self-supervised approach is chosen.

\end{abstract}

\section{Introduction}
\label{sec:intro}

Supervised learning has been extremely successful in recent years in machine learning, except the huge quantity of labeled data needed causes the major problem.
On the other hand, self-supervised learning tries to train the model using only unlabeled data, such as reconstructing the original data from some transformed representations or leveraging some parts of data to predict the other parts, therefore becomes highly attractive.
In natural language processing (NLP) \cite{devlin2018bert,peters2018deep,radford2018improving}, BERT \cite{devlin2018bert} learned powerful representations by self-supervised pre-training to encode contextual information.
In computer vision (CV) \cite{chen2020simple,chen2020big,oord2018representation,bachman2019learning,henaff2019data}, SimCLRv2 \cite{chen2020big} outperformed the previous state-of-the-art on ImageNet by self-supervised pre-training.
Examples in NLP and CV have shown self-supervised pre-trained models are more label-efficient than previous semi-supervised training methods.
In the speech processing area, self-supervised learning also showed great advantages when labeled data are limited~\cite{oord2018representation,schneider2019wav2vec,baevski2020wav2vec,chung2016audio,chung2019unsupervised,liu2020mockingjay,liu2020tera}. 
CPC \cite{oord2018representation} and APC \cite{chung2019unsupervised} learned to extract useful representations for speech using a probabilistic contrastive loss to capture information for predicting future samples.
Wav2vec \cite{schneider2019wav2vec} benefited from the idea of CPC and outperformed the state-of-the-art in character-based ASR with representations learned from 1000 hours of unlabeled speech.
Wav2vec 2.0 \cite{baevski2020wav2vec} further showed that 10 minutes of labeled data were enough for training an ASR system with 53k hours of unlabeled data.
TERA \cite{liu2020tera} pre-trained a Transformer model with a BERT-like objective. The learned representations were shown to be robust for a wide range of downstream tasks. The model could even outperform supervised learning when fine-tuned with only 0.1\% of labeled data.

On the other hand, speech separation has long been a fundamental problem towards robust speech processing under the real-world acoustic environment, in which the considered speech signal is inevitably disturbed by some additional signals produced by other speakers.
In general, deep learning techniques for single-channel speech separation can be divided into two categories: time-frequency (T-F) domain methods and end-to-end time-domain approaches.
Based on T-F features obtained with short-time Fourier transform (STFT), T-F domain methods separate the T-F features for each source and then reconstruct the source waveforms by inverse STFT \cite{hsu2009improvement,huang2014deep,bruna2015source,hershey2016deep,chen2017deep}.
Time-domain approaches then directly process the mixture waveform using an encode-decoder framework, and this line of research has achieved significant progress in recent years \cite{luo2019conv,luo2020dual,chen2020dual,lam2021sandglasset}.
But both the T-F domain and time-domain approaches suffer from the label ambiguity problem when evaluating the reconstruction errors by matching the ground truths with the estimated signals.
Permutation-invariant training (PIT) \cite{yu2017permutation} has been very useful to handle this problem by dynamically choosing the best label assignment each time.
However, the very unstable label assignment during the early training stage in PIT was shown to lead to slower convergence and lower performance \cite{yang2020interrupted}.

In this paper, we made the following contributions:
\begin{itemize}
\item We point out the self-supervised pre-training is also extremely helpful to speech separation.
\item We show the self-supervised pre-training can effectively stabilize the label assignment in PIT during training speech separation models, and the significantly reduced label assignment switching during training directly lead to faster convergence and improved performance.
\item The proposed approach is shown to be equally useful to all different separation models over different datasets, because PIT has been widely used across almost all speech separation tasks.
\end{itemize}

\section{Label ambiguity problem and permutation invariant training (PIT)}
\label{sec:problem}

\subsection{Label ambiguity problem}
In single-channel speech separation, several speech signals are mixed: $y = \sum^N_{n=1} x_n$, where $N$ is the number of sources; the goal is to extract all individual speech signals ${\{x_n\}}^N_{n=1}$ from the mixed signal $y$.
For simplicity, we consider two sources only, $y = x_1 + x_2$, and employ a model with two outputs, $o_1$ and $o_2$.
There exist two possible label assignments: (1) $o_1$ regresses to $x_1$ and $o_2$ regresses to $x_2$, or (2) $o_1$ regresses to $x_2$ and $o_2$ regresses to $x_1$.
These two label assignments lead to two different loss functions to be used in model training.
There are $N!$ possible label assignments for $N \ge 2$.
Incorrect label assignments naturally force the separation model to be updated to wrong direction, or even possibly destroy what has been learned before.

\subsection{PIT and label assignment switching problem}
Permutation invariant training (PIT) \cite{yu2017permutation} was proposed to solve the above problem.
Every time when the model parameters are to be updated, all possible label assignments as mentioned above are used to calculate the regression loss, and the one with minimum loss is chosen to update the model.
Although such a dynamic label selection principle sounds reasonable, the selected labels can be very different for different training epochs giving a very rugged training path.
A soft version of PIT was proposed to relax the label assignment switching problem between epochs \cite{yousefi2019probabilistic}, but restricted to those with $L2$-based objective functions only.
A cascaded training strategy was then proposed \cite{yang2020interrupted}, in which a good label assignment was first obtained with PIT, based on which the model parameters were better updated, to be used as a good initialization for the third stage of PIT training.
This approach properly reduced the assignment switching during training, but made the training time several times longer compared to the original PIT.

\section{Proposed training strategies}
\label{sec:pretrain}

\begin{figure*}[t]
    \centering
    \includegraphics[width=0.93\textwidth]{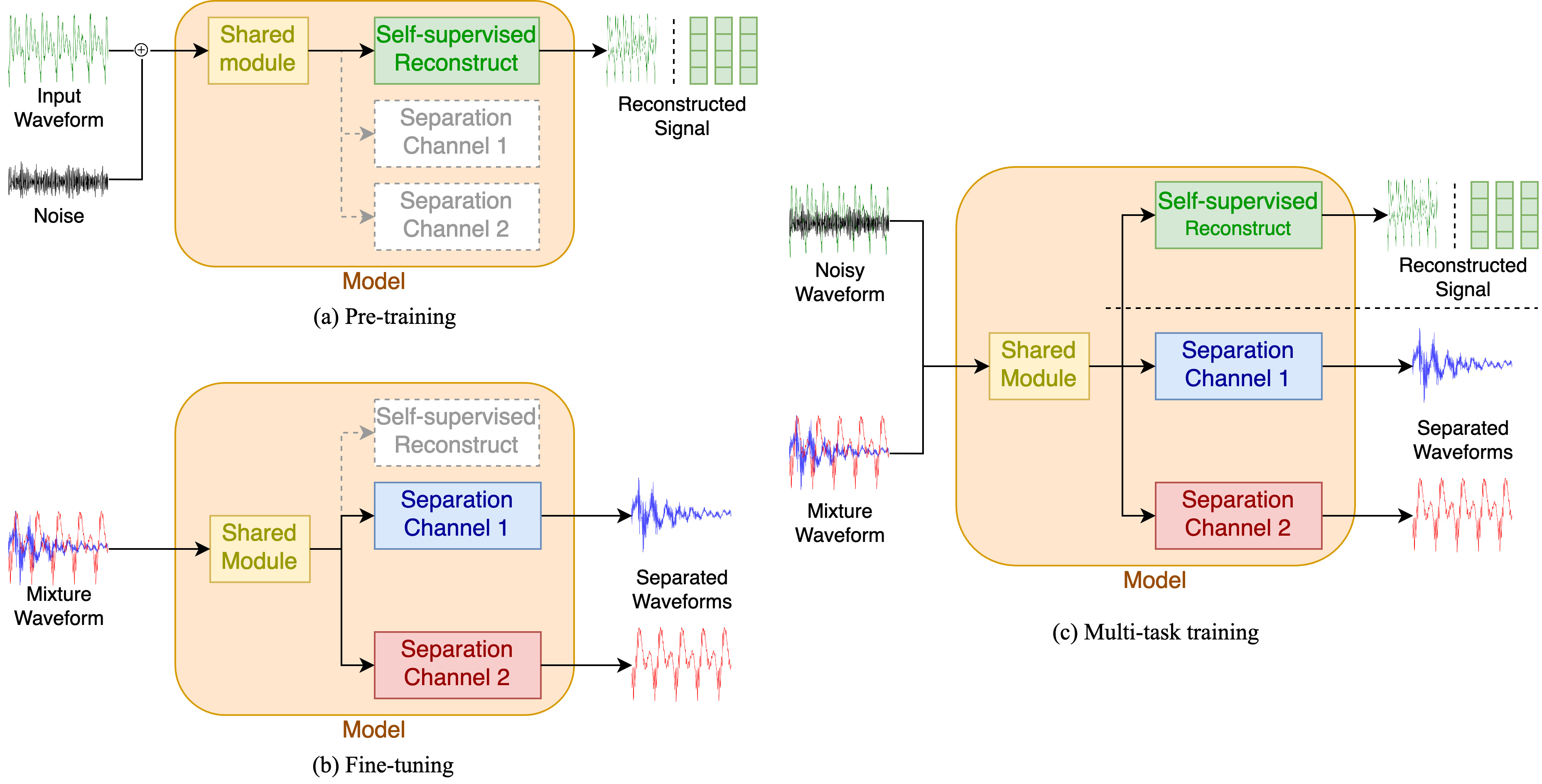}
    \caption{The flowchart for the proposed training framework: (a) pre-training, (b) fine-tuning after pre-training, and (c) multi-task training for comparison. Gray blocks indicate the corresponding parts are not used during training.}
    \label{fig:flowchart}
\end{figure*}

Considering the unstable label assignment problem during training as mentioned above, plus the fact that self-supervised pre-training was shown to be able to assist the model to learn structural information from large-scale unlabeled data and benefit in boosting the following training procedures \cite{devlin2018bert,peters2018deep,radford2018improving,chen2020simple,chen2020big,oord2018representation,bachman2019learning,henaff2019data}, we propose a self-supervised pre-training and fine-tuning framework as below.

\subsection{Pre-train}
\label{ssec:pretrain}
In this work, we consider three different self-supervised approaches for pre-training here: speech enhancement (SE), Masked Acoustic Model with Alteration (MAMA) used in TERA \cite{liu2020tera}, and continuous contrastive task (CC) used in wav2vec 2.0 \cite{baevski2020wav2vec}.
Speech enhancement (SE) simply tries to  reconstruct the original signal when noise is added to the input.
MAMA is a masked reconstruction task, where the input audio is disturbed by noise with some parts randomly picked up and masked, and the model is required to reconstruct the clean audio of the masked parts.
CC is a contrastive task; we mask the spans of the input audio features, and the model is trained to predict the masked spans of features correctly.
Fig. \ref{fig:flowchart}(a) (colored part) shows the flowchart of pre-training, where the input signal is probably mixed with random noise and masked, and the model is to reconstruct the original clean source.

\subsection{Fine-tune}
\label{ssec:finetune}
After pre-training, the model is then fine-tuned with the normal separation training objective to produce the desired individual signals.
All model parameters for fine-tuning are loaded from the pre-trained model as long as available, but the parameters used to generate specific output channels are re-initialized, as shown in Fig. \ref{fig:flowchart}(b).
PIT is performed as usual.

\subsection{Multi-task learning}
\label{ssec:multi-task}
To verify that whether the proposed framework really benefits from the "pre-train then fine-tune" procedure, jointly learning from the self-supervised training plus separation in a multi-task learning framework is also tested as a baseline for comparison, as in Fig. \ref{fig:flowchart}(c). 


\section{Experimental setup}
\label{sec:exp-setup}

\subsection{Dataset}
\label{ssec:dataset}

In this work, speech separation was trained and evaluated on WSJ0-2mix \cite{hershey2016deep} and Libri2Mix \cite{cosentino2020librimix}  \texttt{train-100} set, and self-supervised approaches were trained using Libri1Mix \cite{cosentino2020librimix} \texttt{train-360} set \cite{cosentino2020librimix}.
The WSJ0-2mix dataset was derived from the WSJ0 data corpus \cite{garofolo1993csr}. The training and validation data contained two-speaker mixtures generated by randomly selecting utterances from different speakers in the WSJ0 \texttt{si\_tr\_s} set, and the test set was similarly generated using utterances from unseen speakers in WSJ0 \texttt{si\_dt\_05} and \texttt{si\_et\_05} set.
Libri2Mix is created based on the Librispeech dataset \cite{panayotov2015librispeech} with a similar generating procedure as WSJ0-2mix. 
Libri2Mix \texttt{train-100} set used speakers randomly selected from the \texttt{train-clean-100} set of Librispeech, while the dev and test set used the utterances from unseen speakers in the \texttt{dev} and \texttt{test} sets of Librispeech respectively.
Libri1Mix \texttt{train-360} dataset was created with the same settings as Libri2Mix, while only one speaker was randomly selected from the \texttt{train-clean-360} set of Librispeech, and mixed with a random noise sampled from WHAM! \cite{wichern2019wham}. The speaker groups of Libri1Mix \texttt{train-360} and Libri2Mix \texttt{train-100} set were disjoint.

\subsection{Implementation details}
\label{ssec:detail}

The proposed self-supervised pre-training can be used with any separation model, and the study here was mainly focused on the effectiveness of pre-training.
In this work, we choose Conv-TasNet \cite{luo2019conv} as our main baseline model, DPRNN \cite{luo2020dual} and DPTNet \cite{chen2020dual} were also used in later experiments.
All experiments were implemented with Asteroid \cite{pariente2020asteroid}, and the detailed training configurations are in the repository\footnote{\url{https://github.com/SungFeng-Huang/SSL-pretraining-separation/tree/main/local}}.

The model was trained with three different strategies for comparison in our experiments: from scratch, pre-trained then fine-tune (PT-FT), and multi-task training.
We purposely let the three strategies have the same number of update steps in training the separation task for fairness.
Separation performance was evaluated in scale-invariant signal-to-noise ratio improvement (SI-SNRi) \cite{luo2018speaker} and signal-to-distortion ratio improvement (SDRi) \cite{vincent2006performance}.

\section{Experimental results}
\label{sec:exp}

\subsection{Comparison between the self-supervised pre-training tasks}
\label{ssec:pretraining-tasks-result}

\begin{table}[t]
    \centering
    \caption{Comparison between different self-supervised pre-training approaches when fine-tuned with Conv-TasNet in SI-SNRi and SDRi. The first row is for training from scratch.}
    \label{table:result2}
    \begin{tabular}{c|c|c} 
        \toprule
        Pre-training task & SI-SNRi (dB) & SDRi (dB) \\
        \midrule
        -- & 15.6 & 15.8 \\ 
        \midrule
        SE & \textbf{16.3} & \textbf{16.5} \\
        MAMA \cite{liu2020tera} & 16.2 & \textbf{16.5}\\
        CC \cite{baevski2020wav2vec} & 15.5 & 15.8 \\
        \bottomrule
    \end{tabular}
\vspace{-0.3cm}
\end{table}

We first wished to find out which self-supervise pre-training approach was more helpful to the separation task.
Speech enhancement (SE), Masked Acoustic Model with Alternation (MAMA) and continuous contrastive task (CC) as described in Section \ref{ssec:pretrain} were tested.
We used Conv-TasNet as our separation model.
After pre-trained with SE, MAMA and CC respectively, we fine-tuned the obtained models for 100 epochs for the speech separation task.
The pre-training tasks were all trained on Libri1Mix \texttt{train-360} set, and the fine-tuning task was trained on WSJ0-2mix.
The results listed in Table \ref{table:result2} showed that both SE and MAMA led to significant improvement, but not CC.
Note that both SE and MAMA had input speech disturbed by noise, while the model was to reconstruct the whole utterance (SE) or only the masked parts (MAMA), as mentioned in Section \ref{ssec:pretrain}.
So we may conclude that approaches trying to reconstruct the clean input speech from the noisy and/or masked one are probably more effective for pre-training speech separation tasks.
A possible explanation may be here SE and MAMA already learned to extract from disturbed signals the information about each individual speaker, so all the following Conv-TasNet model needed to learn is to separate the extracted information into two channels, therefore the learning process was more stable and efficient.
This is why in the following tests we only used speech enhancement (SE) for self-supervised pre-training.

\subsection{Effectiveness of pre-training and fine-tuning (PT-FT)}
\label{ssec:results-wsj-libri}
\begin{figure*}[hbt!]
    \centering
    \includegraphics[width=0.97\textwidth]{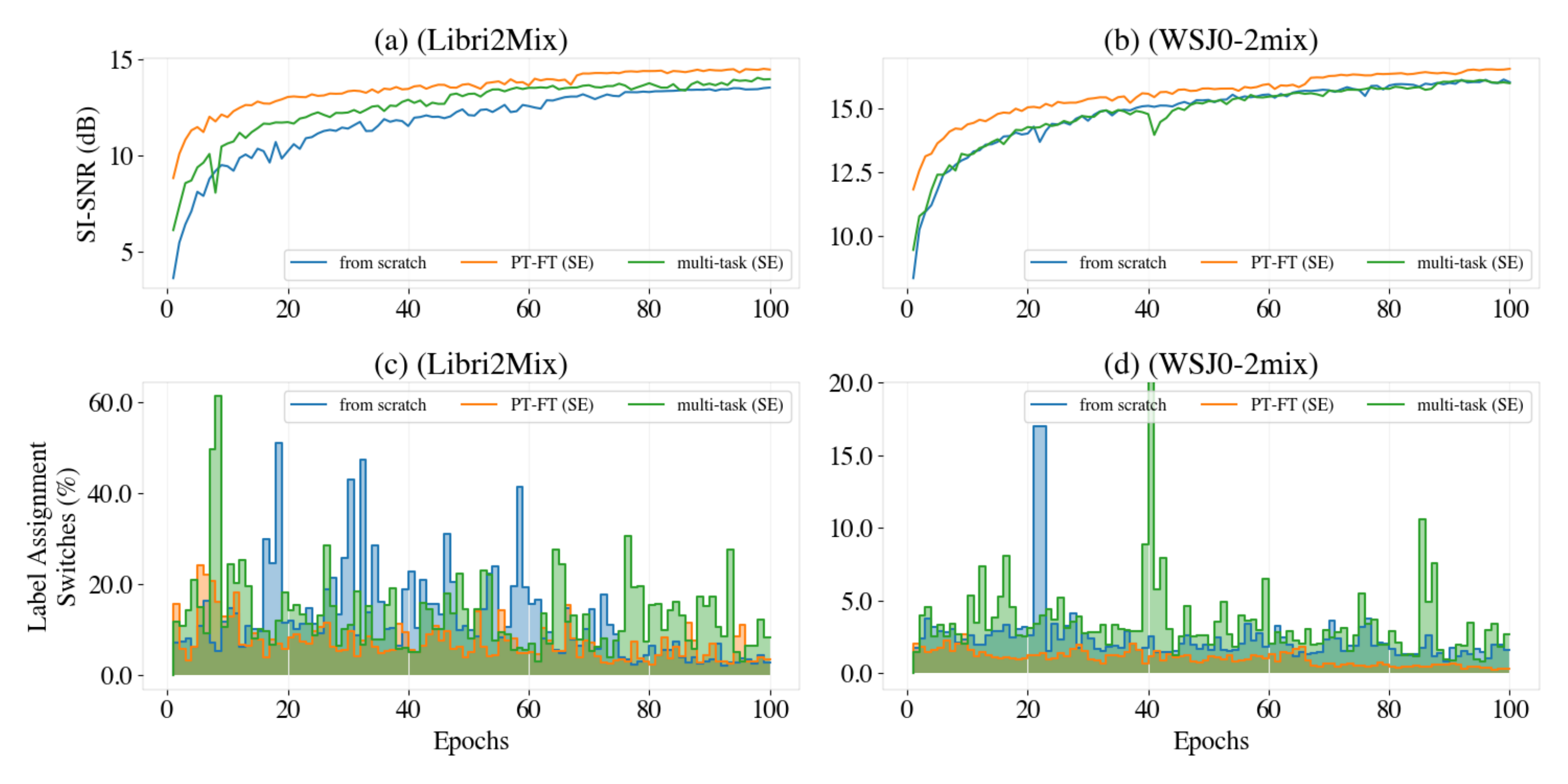}
    \caption{
        (a)(b) validation SI-SNR (dB) and (c)(d) percentage of label assignment switches in total training data (\%) at each epoch on two datasets Libri2Mix and WSJ0-2mix respectively.
        In (d), the green bars reach 89\% at around epoch 40.}
    \label{fig:result}
    \vspace{-0.5cm}
\end{figure*}

\begin{table}[t]
    \centering
    \caption{
        Comparison between different training strategies for Conv-TasNet on two datasets (WSJ0-2mix and Libri2Mix) in SI-SNRi (dB) and SDRi (dB).
        \textsc{\bf PT-FT}: pre-trained then fine-tune.
        \textsc{\bf SE}: with speech separation for self-supervised.
        Number in the parentheses are the improvements over "from scratch".
    }
    \label{table:result}
    \begin{tabular}{c|c|c|c} 
        \toprule
        Corpus & Training strategy & SI-SNRi & SDRi \\
        \midrule
        \multirow{3}{5em}{\centering Libri2Mix} & from scratch & 13.2 & 13.6 \\ 
        & PT-FT (SE) & \textbf{14.1 (0.9)} & \textbf{14.6 (1.0)} \\
        & multi-task (SE) & 13.7 (0.5) & 14.1 (0.5) \\
        \midrule
        \multirow{3}{5em}{\centering WSJ0-2mix} & from scratch & 15.6 & 15.8 \\ 
        & PT-FT (SE)  & \textbf{16.3 (0.7)} & \textbf{16.5 (0.7)} \\
        & multi-task (SE) & 15.7 (0.1) & 16.0 (0.2) \\
        \bottomrule
    \end{tabular}
\vspace{-0.3cm}
\end{table}

Table \ref{table:result} shows the results of three different training strategies: from scratch, pre-training and fine-tuning (PT-FT) and multi-task, with the latter two using speech enhancement (SE) for self-supervised learning, all trained with Conv-TasNet as the main separation model.
As shown, both pre-training and multi-task learning improved the separation model on both WSJ0-2mix and Libri2Mix, while pre-training improved more significantly (0.7 - 1.0 dB improvements for PT-FT (SE) compared to "from scratch" v.s. 0.1 - 0.5 dB for multi-task (SE)).
A good explanation for this is that, as mentioned above, the pre-trained model already learned to extract from disturbed signals the information about the individual target speakers, so the following separation model could focus on the construction of the two masks, for the two sources.
In contrast, for multi-task learning, the two different tasks of speech enhancement and speech separation were learned jointly, while sharing the knowledge learned for the two very different tasks may not be easy.
This further showed the effectiveness of learning the two different tasks sequentially instead of jointly (self-supervised for enhancement then fine-tuning for separation).
Also noted that since the corpus used to train speech enhancement was Libri1Mix \texttt{train-360} set, which was closer to Libri2Mix \texttt{train-100} set but farther from WSJ0-2mix, which may be the simple reason why the results in Table \ref{table:result} on Libri2Mix (upper half) showed more improvements than those on WSJ0-2mix (lower half).

Validation SI-SNR results during training are reported in Figure \ref{fig:result}(a)(b) for Libri2Mix and WSJ0-2mix respectively.
As shown in the figure, improvements for multi-task learning gradually decreased while training on Libri2Mix and were nearly hard to see on WSJ0-2mix.
The proposed pre-trained model led the baselines all the way and achieved the final result of the baselines in only 37 epochs for Libri2Mix and 66 epochs for WSJ0-2mix, which are about one-third to two-thirds of the baseline training epochs.
Figure \ref{fig:result}(c)(d) show the percentage of label assignment switches in total training data.
Here we can see only the proposed pre-training with speech enhancement (orange bars) significantly reduced label assignment switches, while multi-task learning (green bars) not only failed to reduce the label assignment switches but sometimes increased them.
Moreover, training from scratch (blue bars) and multi-task learning (green bars) sometimes got very high switching percentages (e.g., roughly 15\% - 35 \% of the label assignments were often switched over for both training from scratch and multi-task training in Figure \ref{fig:result}(c), and most label assignments were switched at epoch 40 for multi-task training in Figure \ref{fig:result}(d).


\subsection{More Separation models tested on WSJ0-2mix}
\label{ssec:wsj0-result}

\newcolumntype{s}{>{\columncolor{blue!30!}} c}

\begin{table}[t]
    \centering
    \caption{
        Different separation models on WSJ0-2mix in SI-SNRi (dB) and SDRi (dB).
        \textsc{\bf BS}: batch size.
        \textsc{\bf L}: utterance length (sec).
        \textsc{\bf PT}: pre-training, "--" means training from scratch.
        The first rows (a)(d)(g)(l) for each model are the reported results from original papers. The blank indicates unknown.
        *Row (g) are actually SI-SNR and SDR.}
    \label{table:wsj0-result}
    \begin{tabular}{c|s|c|c|c|c|c} 
        \toprule
        \rowcolor{white}
        & Model & BS & L & PT & SI-SNRi & SDRi\\
        \midrule
        
        \rowcolor{red!30!}
        (a) & \cellcolor{blue!30!}
            &       & 4     & --    & 15.3          & 15.6 \\
        (b) & 
            & 24    & 3     & --    & 15.6          & 15.8 \\ 
        (c) & \multirow{-3}{6em}{\centering (I)\\Conv-TasNet}
            & 24    & 3     & SE    & \textbf{16.3} & \textbf{16.5} \\
        \hlinewd{0.5pt}

        \rowcolor{red!30!}
        (d) & \cellcolor{blue!30!}
            &       & 4     & --    & 18.8          & 19.0 \\ 
        (e) &
            & 24    & 2     & --    & 17.0          & 17.3 \\ 
        (f) & \multirow{-3}{5em}{\centering (II)\\DPRNN}
            & 24    & 2     & SE    & \textbf{18.6} & \textbf{18.9} \\
        \hlinewd{0.5pt}
        
        \rowcolor{red!30!}
        (g) & \cellcolor{blue!30!}
            &       & 4     & --    & 20.2*         & 20.6* \\ 
        (h) & 
            & 4     & 4     & --    & 20.4          & 20.6 \\
        (i) & 
            & 4     & 4     & SE    & \textbf{20.8} & \textbf{21.0} \\
        \cline{1-1}\cline{3-7}
        (j) &
            & 1     & 4     & --    & 20.7          & 20.9 \\ 
        (k) &\multirow{-5}{5em}{\centering (III)DPTNet}
            & 1     & 4     & SE    & \textbf{21.3} & \textbf{21.5} \\
        \hlinewd{0.5pt}
        
        \rowcolor{red!30!}
        (l) & \cellcolor{blue!30!} (IV) Sandglasset
            &       & 4     & --    & 21.0          & 21.2\\
        \bottomrule
    \end{tabular}
\vspace{-0.3cm}
\end{table}

More test results on different separation models with different batch sizes (BS), utterance length (L), with pre-training (PT) or from scratch are listed in Table \ref{table:wsj0-result}, all trained and evaluated on WSJ0-2mix.
The first rows (a)(d)(g)(l) for each model are those reported in their original papers.
For speeding up the experiments, Conv-TasNet and DPRNN (Sec. (I)(II) in Table \ref{table:wsj0-result}) were trained with shorter utterance length (3 or 2 sec) and a larger batch size (24) with 200 epochs, which caused the slightly worse DPRNN results than those previously reported \cite{luo2020dual}.
In addition to those for Conv-TasNet discussed previously, the pre-trained DPRNN (Sec. (II)) was improved significantly, even achieving comparable performance as the reported one (rows (f) v.s. (d)(e)), although with worse performance from scratch due to the hyper-parameters.
DPTNet (Sec. (III)) was trained with batch size 1 and 4 with 100 epochs to speed up the training process.
Setting batch size 4 instead of 1 gave 0.3 dB worse performance (rows (h) v.s. (j)).
Nevertheless, the pre-trained DPTNet made up the gap, even doing slightly better (rows (i) v.s. (j)).
Compared to the current state-of-the-art (Sandglasset \cite{lam2021sandglasset}), the pre-trained DPTNet with a batch size 1 actually achieved the new state-of-the-art ((k) v.s. (l)).

\section{Conclusion}
\label{sec:print}

In this paper, we propose to use self-supervised pre-training to stabilize the label assignment for speech separation.
We show that pre-training with speech enhancement offers better training and consistently improves the separation performance across all different separation model architectures over two different datasets.


\section{Acknowledgements}

We thank to professor Lin-shan Lee for the guidance and National Center for High-performance Computing (NCHC) of National Applied Research Laboratories (NARLabs) in Taiwan for providing computational and storage resources.


\bibliographystyle{IEEEtran}


\end{document}